\documentclass[
superscriptaddress,
twocolumn,
 amsmath,amssymb,
 aps]{revtex4-2}

\usepackage{graphicx}% Include figure files
\usepackage{dcolumn}% Align table columns on decimal point
\usepackage{bm}% bold math
\usepackage{color}
\usepackage{xcolor}
\usepackage{hyperref}% add hypertext capabilities
\usepackage{tabularx}
\usepackage{placeins}

\newcommand{\birte}[1]{\textcolor{black}{#1}}

\newcommand{\red}[1]{\textcolor{black}{#1}}

\begin{document}

\title{Spatiotemporal Control of Charge +1 Topological Defects in Polar Active Matter}

\author{Birte C. Geerds}
\affiliation{Faculty of Physics, TU Dresden, 01069 Dresden, Germany}
\affiliation{Department of Theoretical Physics, University of Geneva, 1211 Geneva, Switzerland}
\author{Abhinav Singh}
\affiliation{Dresden University of Technology, Faculty of Computer Science, 01187 Dresden, Germany}
\affiliation{Max Planck Institute of Molecular Cell Biology and Genetics, 01307 Dresden, Germany}
\affiliation{Center for Systems Biology Dresden, 01307 Dresden, Germany}
\author{Mathieu Dedenon}
\affiliation{Department of Biochemistry, University of Geneva, 1211 Geneva, Switzerland}
\affiliation{Department of Theoretical Physics, University of Geneva, 1211 Geneva, Switzerland}
\author{Daniel J.~G.~Pearce}
\affiliation{Department of Theoretical Physics, University of Geneva, 1211 Geneva, Switzerland}
\author{Frank J\"ulicher}
\affiliation{Max Planck Institute for the Physics of Complex Systems, 01187 Dresden, Germany}
\affiliation{Faculty of Physics, TU Dresden, 01069 Dresden, Germany}
\affiliation{Center for Systems Biology Dresden, 01307 Dresden, Germany}
\affiliation{Cluster of Excellence ``Physics of Life'', TU Dresden, 01307 Dresden, Germany}
\author{Ivo F.~Sbalzarini}
\affiliation{Faculty of Computer Science, TU Dresden, 01187 Dresden, Germany}
\affiliation{Max Planck Institute of Molecular Cell Biology and Genetics, 01307 Dresden, Germany}
\affiliation{Center for Systems Biology Dresden, 01307 Dresden, Germany}
\affiliation{Cluster of Excellence ``Physics of Life'', TU Dresden, 01307 Dresden, Germany}
\author{Karsten Kruse}
\affiliation{Department of Biochemistry, University of Geneva, 1211 Geneva, Switzerland}
\affiliation{Department of Theoretical Physics, University of Geneva, 1211 Geneva, Switzerland}

\date{\today}

\begin{abstract}
Topological defects are a conspicuous feature of active liquid crystals that have been associated with important morphogenetic transitions in organismal development. Robust development thus requires a tight control of the motion and placement of topological defects. 
In this manuscript, we study a mechanism to control +1 topological defects in an active polar fluid confined to a disk. If activity is localized in an annulus within the disk, the defect moves on a circular trajectory around the center of the disk.
Using an \textit{ansatz} for the polar field, we determine the dependence of the angular speed and the circle radius on the boundary orientation of the polar field and the active annulus.
Using a proportional integral controller, we guide the defect along complex trajectories by changing the active annulus size and the boundary orientation.
\end{abstract}

\maketitle

\section{Introduction}

Active stress generated by the transformation of chemical energy into mechanical work is at the base of fascinating autonomous movements seen in biological systems~\cite{Prost:2015ev,balasubramaniamActiveNematicsScales2022}. At the molecular level, it results from the interaction of molecular motors with filamentous protein assemblies. Due to their molecular form, active stress in living matter is often anisotropic. The principal direction of the stress typically coincides with orientational order exhibited by the biological material. Macroscopic orientational order can result, for example, from the alignment of filaments~\cite{Sanchez:2012gt} or of elongated cells~\cite{Gruler:1999bt,Duclos:2014bs}. If individual filaments or cells are polarized, their alignment can result in polar or nematic orientational order.

Topological point defects are increasingly recognized to play a central role in organizing mechanical stress in biological systems. At such  defects, the orientational order field is ill defined. They are characterized by their topological charge, which can be either half-integer or integer for nematic order, whereas only integer charges occur for polar order. Defects have been reported to determine the position of cell death and extrusion~\cite{Saw:2017gn}, direct the formation of cell mounds~\cite{Kawaguchi:2017em,Guillamat.2022, EndresenMurali.2026}, and are essential for surface deformations that accompany the formation of mouth, foot, and tentacles during regeneration of the freshwater polyp \textit{Hydra vulgaris}~\cite{Maroudas-Sacks.2021,Ravichandran.2025}. Activity-induced surface deformations orchestrated by topological defects also hold promises in the design of soft robots~\cite{Webster-Wood_2023}.

Topological defects of charge +1 play a particular role in surface deformation. These defects are unique in that they have continuous rotational symmetry. A homogeneous local rotation of the orientation field, referred to as a change in phase, causes a +1 defect to transition between aster, spiral and vortex. For any other defect, a change in phase is simply associated with a global rotation around the defect center. Active defects of charge +1 can deform surfaces into different shapes ranging from domes to saddles depending on the phase of the defect~\cite{Pearce.2020}. Thus, +1 defects offer the possibility to control shapes by changing their phase. Phase changes can be performed in a restricted region around the defect, whereas the alternative, changing the topological charge of a defect, requires global changes of the orientational order.

It is thus desirable for biological and engineered systems to be able to control the arrangement of topological defects. Several proposals have been made to achieve spatio-temporal control of defect patterns in active nematics~\cite{Norton.2020,Zhang.2021, Tang.2021, Ruske.2022, ronning_defect_2023, Shankar.2024, floydTailoringInteractionsActive2025}. 
Such ideas have been realized in experiments on reconstituted cytoskeletal motor-filament systems~\cite{Zhang.2021,Nishiyama.2025}. Similar activity patterning techniques have been shown to be applicable to Toner--Tu polar systems~\cite{Ghosh.2024}. Defects of charge +1 are rotationally symmetric and, in contrast to half-integer defects, do not self-propel. Persistent motion of integer defects was achieved using dynamic activity patterns~\cite{Ghosh.2024}. 
In spite of these efforts, achieving controlled positioning and persistent motion of +1 defects remains challenging. We explore how symmetry breaking and heterogeneous activity can work together for a precise control of such defects.

In this work, we focus on a +1 topological defect in an active polar fluid confined to a disk. We show how heterogeneous activity induces defect self-propulsion and explain its physical origin. We then leverage this mechanism for spatiotemporal defect control using a closed-loop controller. We achieve precise radial and azimuthal motion and discuss potential applications.

\section{Defect dynamics on a disk}

In the following, we briefly recall the hydrodynamic equations of an active polar fluid, where the magnitude of the polar order is not fixed. We then consider the steady state of such a fluid confined to a disk, where the boundary conditions impose a total defect charge of +1. Whereas the defect co-localizes with the disk center for homogeneous activity, it can circulate around the center if the fluid is passivated in the disk's central region.

\subsection{Hydrodynamics of an active polar fluid}\label{subsec:hydrodynamics_of_an_active_polar_fluid}

We investigate an incompressible 2D polar active fluid confined to a disk using generalized hydrodynamics~\cite{Kruse:2004il, Kruse:2005fy, Fuerthauer_2012}. The polar field is described by the vector $\mathbf{p}$, which is associated with the free energy $F=\int f\,dA$, where
\begin{align}\label{eq:elastic_energy}
    f &= \frac{K_S}{2}(\nabla\cdot\mathbf{p})^2 + \frac{K_B}{2}(\nabla\times\mathbf{p})^2
    +\frac{\chi}{2}\left(\frac{|\mathbf{p}|^4}{2} - |\mathbf{p}|^2 \right)\, .
\end{align}
The first two terms describe how the free energy depends on splay and bend distortions of $\mathbf{p}$ with respective elastic coefficients $K_S$ and $K_B$. The parameter $\chi$ controls the magnitude of a soft constraint for $|\mathbf{p}|^2 = 1$. We will refer to $|\mathbf{p}|$ as the polar order parameter.

The dynamics of $\mathbf{p}$ are governed by 
\begin{equation}
\label{eq:time_evolution_p}
    \frac{D}{Dt}p_{\alpha} = \frac{1}{\gamma}h_{\alpha} + \lambda p_{\alpha} \Delta \mu - \nu  u_{\alpha \beta}p_{\beta} \;.
\end{equation}
In this expression, Greek indices go from 1 to 2 and  we adopt the convention of summing over repeated indices. The molecular field $h_{\alpha} = -\frac{\delta F}{\delta p_\alpha}$ describes the response of $p_\alpha$ to deviations from its equilibrium configuration and the parameter $\gamma$ represents rotational viscosity. Activity results from chemical processes driven by the chemical potential difference $\Delta \mu$. It couples directly to the dynamics of the polarization via the phenomenological coefficient $\lambda$. Possible molecular mechanisms leading to such a term include cytoskeletal filament disassembly, which involves hydrolysis of nucleotide tri-phosphates, and upstream activation or deactivation of regulators of filament assembly, for example, by small GTPases. Note that, for $\lambda\Delta\mu<0$, activity tends to decrease polar order.
The parameter $\nu$ is the flow alignment coefficient, which couples the polarization angle to the strain-rate tensor 
$u_{\alpha\beta}=\frac{1}{2}(\partial_{\alpha}v_{\beta}+\partial_{\beta}v_{\alpha})$,
where $\mathbf{v}$ is the velocity field. The co-rotational material derivative is given by  
$Dp_{\alpha}/Dt = \partial_t p_{\alpha} + v_{\gamma} \partial_{\gamma}p_{\alpha} + \omega_{\alpha \beta} p_{\beta}$ with the vorticity tensor  
$\omega_{\alpha\beta}=\frac{1}{2}(\partial_{\alpha}v_{\beta}-\partial_{\beta}v_{\alpha})$.

We consider flows with low Reynolds number and assume that there are no external forces. Force balance then reads \begin{align}
\partial_\beta\sigma_{\alpha\beta} = 0,
\label{eq:forceBalance}
\end{align}
for $\alpha=1,2$, where the total stress tensor can be decomposed into a symmetric and an antisymmetric part $\sigma_{\alpha\beta} = \sigma_{\alpha\beta}^\mathrm{sym} + \sigma_{\alpha\beta}^\mathrm{ant} $. The symmetric part of the total stress has components from the Ericksen, viscous, (symmetric) reactive and active stress: $\sigma_{\alpha\beta}^\mathrm{sym} = \sigma^{e,\mathrm{sym}}_{\alpha\beta} + \sigma^v_{\alpha\beta} + \sigma^r_{\alpha\beta} + \sigma^a_{\alpha\beta}$. Here, the total Ericksen or elastic stress is given by 
\begin{equation}
    \sigma^e_{\alpha\beta} =  -P\delta_{\alpha\beta} - \frac{\partial f}{\partial(\partial_{\beta} p_{\gamma})} \partial_{\alpha} p_{\gamma} \,,
\end{equation}
where $P$ is the hydrostatic pressure. It is further decomposed into a symmetric and an antisymmetric part. The symmetric part $\sigma^{e,\mathrm{sym}}_{\alpha\beta}=\frac12(\sigma^{e}_{\alpha\beta}+\sigma^{e}_{\beta\alpha})$ generalizes the hydrostatic pressure in the presence of the polarization field $\mathbf{p}$. The total antisymmetric stress reads
\begin{equation}
    \sigma_{\alpha\beta}^\mathrm{ant} =  \left(p_{\alpha}h_{\beta}-p_{\beta}h_{\alpha}\right)/2 + \sigma_{\alpha\beta}^{e,\mathrm{ant}} \,,
\end{equation}
where $\sigma_{\alpha\beta}^{e,\mathrm{ant}}=\frac12(\sigma^{e}_{\alpha\beta}-\sigma^{e}_{\beta\alpha})$ is the antisymmetric part of the Ericksen stress. The antisymmetric part of the total stress contributes to the angular momentum balance by mediating exchange between orbital and internal angular momentum. The remaining terms for viscous, reactive and active stress components, respectively are given by
\begin{align}
    \sigma^v_{\alpha\beta} &= 2\eta u_{\alpha\beta}\\
    \sigma^r_{\alpha\beta} &= \frac{\nu}{2} \left( p_{\alpha} h_{\beta} + p_{\beta} h_{\alpha} - p_{\gamma} h_{\gamma} \delta_{\alpha \beta} \right) \\
    \label{eq:active_stress}
    \sigma^a_{\alpha\beta} &= -\zeta \Delta \mu \left( p_{\alpha} p_{\beta} - \frac{1}{2} p_{\gamma} p_{\gamma} \delta_{\alpha \beta} \right)\, .
\end{align}
Here, $\eta$ the fluid's viscosity, and $\nu$ the flow alignment parameter. The parameter $\zeta$ couples the active chemical processes to the mechanics. The equations are closed by the incompressibility condition $\partial_{\gamma} v_{\gamma} = 0$, which determines $P$.

In  the disk geometry, it is convenient to express the polar field $\mathbf{p}$ by its norm, that is, the polar order parameter $|\mathbf{p}|$, and the orientation angle $\psi$ between $\mathbf{p}$ and the inward radial direction, $\cos(\psi) =-\hat{\mathbf{p}}\cdot\hat{\mathbf{r}}$, where a hat indicates a normalized vector. At the disk boundary $r=r_b$, we implement no-slip conditions for the velocity and impose $\psi=\psi_b$ as well as $|\mathbf{p}|=1$. These boundary conditions impose a topological defect charge $+1$ in the polar field.

In the following, we scale time by $\gamma/\chi$, length by $\sqrt{K_S/\chi}$, and energy by $K_S$. This leaves us with the following dimensionless parameters $\bar{\zeta}=\zeta\Delta\mu/\chi$, $\bar{\lambda}=\gamma\lambda\Delta\mu/\chi$, $\bar{\eta}=\eta/\gamma$, and $\overline{K}_B=K_B/K_S$. Note that both, $\bar{\zeta}$ and $\bar{\lambda}$, depend on the chemical potential difference $\Delta\mu$. In experiments, it is easier to change $\Delta\mu$ than $\zeta$ or $\lambda$. For this reason, $\Delta\mu$ is varied while keeping $\bar{\lambda}/\bar{\zeta}=-7.5$ constant unless stated otherwise , and results are shown in terms of $\bar\lambda$. 

For numerical solutions to the dynamic equations we used a randomized point cloud in space and evaluated the differential operators using Discretization-Corrected Particle Strength Exchange (DC-PSE)~\cite{Schrader.2010}. The scheme was implemented in the OpenFPM library for scalable scientific computing~\cite{Incardona.2018} using a template-expression system for hydrodynamic equations~\cite{Singh.2021}. We solved for Stokes flow with a pressure-correction algorithm to ensure incompressibility~\cite{Singh:2023c} and evolved the system in time using the adaptive Adams--Bashforth--Moulton time-stepping method~\cite{singh_integrating_2026}. See App.~\ref{app:numerics} for details and Tab.~\ref{tab:parameters_constitutiveequ} for  parameter values.

\subsection{Spontaneous defect motion } 

In this section, we study the dynamic equations \eqref{eq:time_evolution_p} and \eqref{eq:forceBalance} for extensile active stress, $\bar{\zeta}\ge0$, which is spatially homogeneous in the region $r\ge r_i$ and vanishes for $r<r_i$. We briefly discuss the case $r_i=0$ and then consider the general case.

\subsubsection{Homogeneous activity}

For homogeneous extensile activity, we obtain rotating spiral defects if the activity exceeds a critical value, App.~\ref{app:homogenousActvity}. This is in agreement with previous reports~\cite{Kruse:2004il}, where the polar order parameter was kept constant throughout. Although in the present system $|\mathbf{p}|$ is not fixed , in thermodynamic equilibrium, the free energy \eqref{eq:elastic_energy} still yields $|\mathbf{p}|=1$ away from defects. Activity, however, modifies the magnitude of the polar order. Specifically, in a steady state with $\mathbf{v}=0$ and $|\mathbf{p}|=const$, the active term in Eq.~\eqref{eq:time_evolution_p} leads to 
\begin{align}
\label{eq:P0}
|\mathbf{p}|^2 &= 1+\bar{\lambda}\equiv P_0^2.
\end{align}
In presence of a flow, the polar order parameter is further affected by flow alignment, see App.~\ref{app:homogenousActvity}. 

\subsubsection{Heterogeneous activity}
\begin{figure}[t]
    \centering
    \includegraphics[width=\columnwidth]{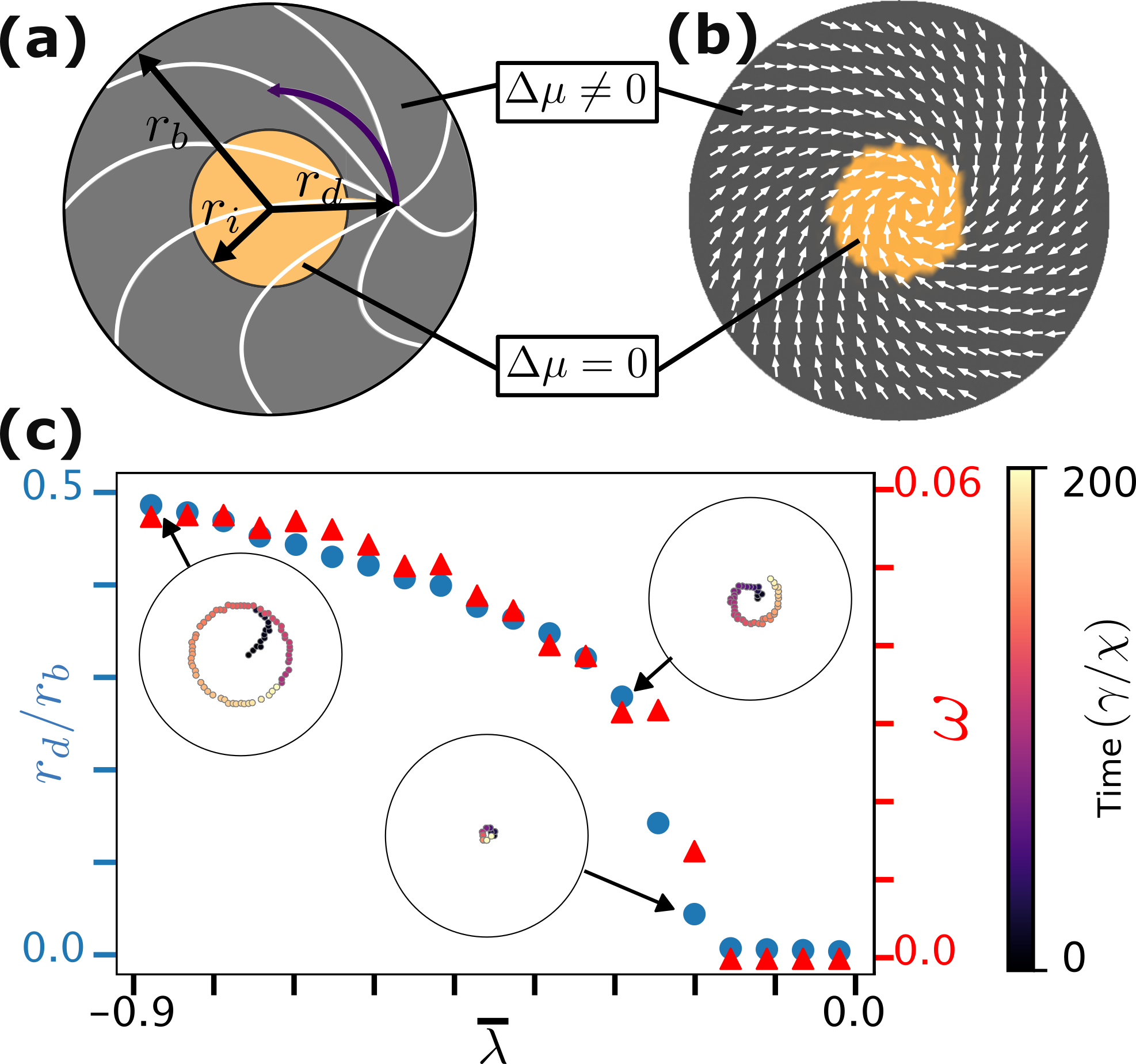}
    \caption{Defect on a disk with heterogeneous activity pattern. \textbf{(a)} Illustration of the system. At the disk boundary, $r=r_b$, we impose $|\mathbf{p}|=1$ and $\psi_b=\pi/4$. Gray: active region, $\bar{\zeta},\bar{\lambda}\neq0$, yellow: passive region, $\bar{\zeta}=\bar{\lambda}=0$. White lines: polarization field lines. Dark purple curved arrow: circular defect path with radius $r_d$.  \textbf{(b)} Initial configuration of the numerical solver. White arrows: polarization field, gray and yellow regions as in (a). \textbf{(c)} Radial defect position $r_d$ (blue dots) and angular defect speed $\omega$ (red triangles)  as a function of activity $\bar{\lambda}$. Insets: Trajectories of the defect for $\bar{\lambda}=-0.857$ (left), $\bar{\lambda}=-0.300$ (right), $\bar{\lambda}=-0.214$ (center). Colors: time along trajectories. Parameters as in Tab.~\ref{tab:parameters_constitutiveequ} and $\bar{\lambda}=-7.5\bar{\zeta}$ and $r_i/r_b=8/25$.}
    \label{fig:heterogeneousActivity}
\end{figure}

We next investigate the dynamics if $r_i>0$, such that the system is passive in a central disk and activity is confined to an annulus extending to the boundary of the disk, Fig.~\ref{fig:heterogeneousActivity}(a). From now on, we use the one-constant approximation, $K_B=K_S$, and set the boundary angle to $\psi_b=\pi/4$. 

For an active stress that is larger than a critical value, $\bar{\lambda}_c<\bar{\lambda}<0$,
the defect stays in the passive region and moves to the center of the disk. The defect adopts a spiral configuration, Fig.~\ref{fig:heterogeneousActivity}(b), and generates a rotationally symmetric flow similar to the homogeneous case, see App.~\ref{app:homogenousActvity}, Fig.~\ref{fig:steadyStateHomogenousActivity}(c,d). For $\bar{\lambda}<\bar{\lambda}_c$ and $\bar{\zeta}>0$, the defect leaves the passive region and settles on a circular trajectory around the center, Fig.~\ref{fig:heterogeneousActivity}(c). The corresponding angular speed $\omega$ and radius $r_d$ increase with the activity. Let us note that the defect remains in the disk center if $\bar{\lambda}\ge0$. 

\begin{figure}[t]
    \centering
    \includegraphics[width=0.8\linewidth]{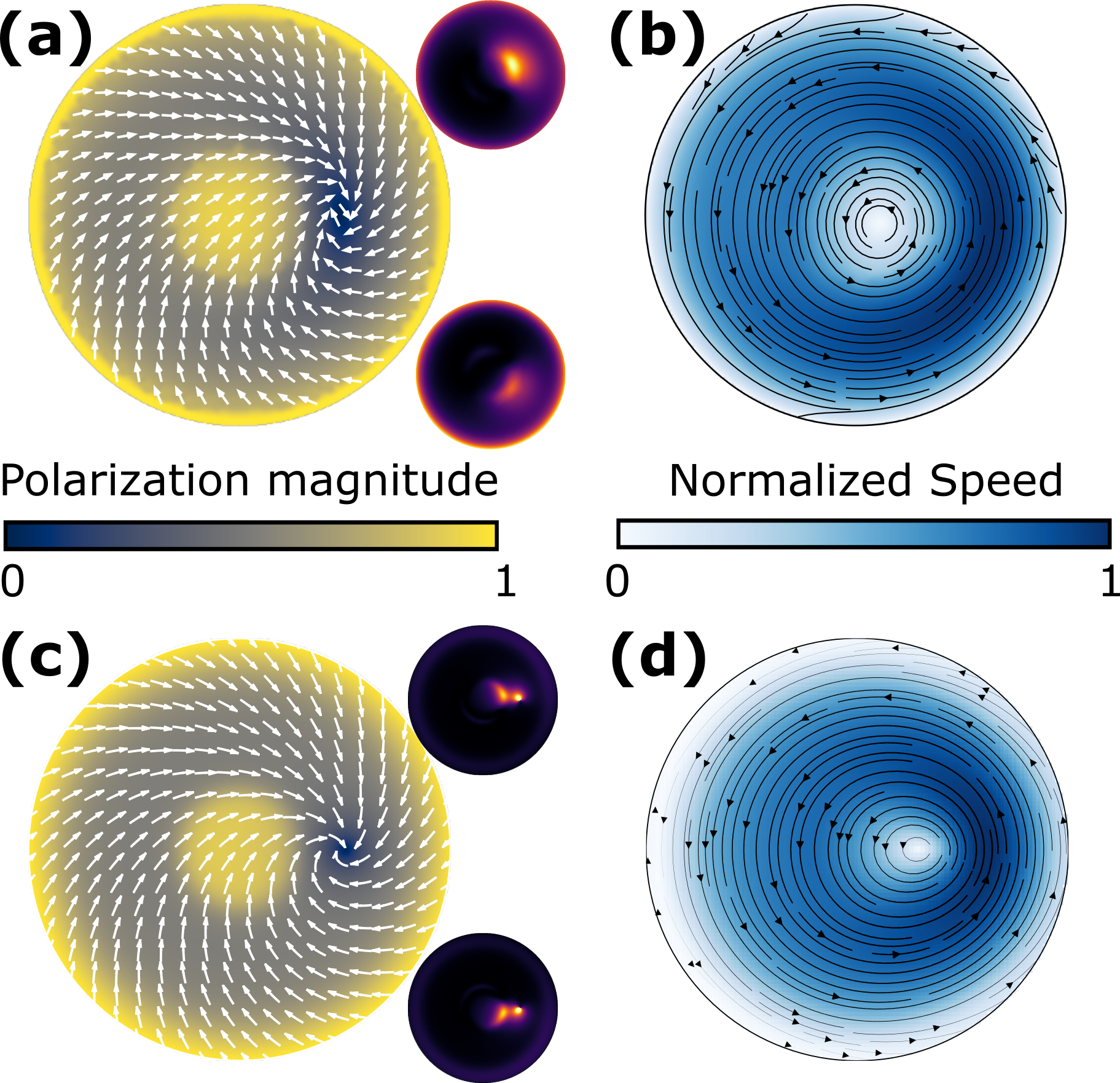}
    \caption{Polarization and flow fields for a defect circulating with constant angular velocity.
    \textbf{(a,b)} Numerical solution of the dynamic equations. (a) White arrows: $\hat{\mathbf{p}}$, colors: $|\mathbf{p}|$. Insets: Splay (top) and bend distortions (bottom) in $\mathbf{p}$. Hot colors indicate stronger deviations from the equilibrium configuration. 
    (b) Black lines: streamlines, colors: speed normalized to its maximum $v_\mathrm{max}=1.14$. Parameters as in Fig.~\ref{fig:heterogeneousActivity}(c) and $\bar{\lambda}=-0.857$.
\textbf{(c,d)} Polarization (c) and flow fields (d) given by the \textit{ansatz} from Sect.~\ref{sec:ansatz}.  
     Parameters: $\bar{p}_0=0.894$, $\Delta P_1=0.383$, $\Delta P_2=0.623$, $r_1=0.038$, $r_2=0.104$, $\lambda_1=0.037$ and $\lambda_2=0.08$. Furthermore, $v_\mathrm{max}=1.07$.
    }
    \label{fig:heterogeneousPolarizationAndFlow}
\end{figure}

For a defect circulating around the disk center, the polarization field is not rotationally symmetric around the defect, even in its vicinity, Fig.~\ref{fig:heterogeneousPolarizationAndFlow}(a). Instead, it shows distinct areas of increased bend or splay, where splay is increased in front of the circulating defect and bend behind, see insets in Fig.~\ref{fig:heterogeneousPolarizationAndFlow}(a). The polar order parameter is larger in the passive than in the active region as expected from Eq.~\eqref{eq:P0}. Moreover, in the passive region, the polarization field approaches a uniform configuration. Still, flows are also present in the passive region, Fig.~\ref{fig:heterogeneousPolarizationAndFlow}(b). As in the case of homogeneous activity, these flows are driven by the active stress $\sigma^a$ and active contributions to the dynamics of $\mathbf{p}$ through $\lambda$.  

The flow speed is largest close to the defect and decreases towards the passive region and also towards the disk boundary, where we impose $\mathbf{v}=0$. At the defect position, the flow field is purely azimuthal and its angular speed similar to that of the defect. 
In the following, we estimate the angular defect velocity $\omega$ through an \textit{ansatz} for the polarization field. 
\section{Control of defect motion}

In the previous section, we found that +1 topological defects circulate around the center of a disk that is divided into a passive central disk and an outer active annulus. 
In the following, we introduce an \textit{ansatz} for the polarization field.  
Based on this \textit{ansatz}, we find that the radius $r_i$ of the passive inner disk and the boundary angle $\psi_b$ can be used as parameters for controlling the defect. Finally, we implement a PI controller to simultaneously modify $\psi_b$ and $r_i$ to set the defect on a prescribed path.

\subsection{An approximate solution for the defect dynamics}
\label{sec:ansatz}

To start, we notice that the orientation of the polar field varies weakly along the radial directions from the defect position to the boundary, Fig.~\ref{fig:heterogeneousPolarizationAndFlow}(a). In our \textit{ansatz} for $\mathbf{p}$, we choose the orientation to be constant along the radial directions from the defect center. As a consequence, the polarization field is fully determined by the boundary angle $\psi_b$ and by the radial defect position $r_d$. 

We use polar coordinates $(r,\theta)$ centered at the disk center and consider a defect located at $(r_d,\theta_d)$. A convenient representation of the polarization field is obtained by introducing a second polar coordinate system $(\rho,\phi)$ centered at the defect. The coordinate systems are oriented such that $\phi=0$ corresponds to the direction $\theta_d$. Without loss of generality, we set $\theta_d=0$. For a point with coordinates $(r,\theta)$ in the disk-centered system and the coordinates $(\rho,\phi)$ in the defect-centered system, we have  
\begin{align}
    \rho &= \sqrt{r^2 - r_d^2\sin^2\phi} - r_d\cos\phi
    \label{eq:rhoOfRPhi}\\
    \tan\theta &= \frac{\rho\sin\phi}{r_d+\rho\cos\phi}.
    \label{eq:thetaOfRPhi}
\end{align}

From the above expressions, we obtain the polarization angle $\Psi(\phi)$ along the direction $\phi$. It is given by the boundary angle $\psi_b$ measured relative to the radial direction from the disk center by 
\begin{align}
    \label{eq:psi}
    \Psi(\phi) = \psi_b+\theta_b(\phi)-\phi.
\end{align}
Here $\theta_b(\phi)$ is given by Eqs.~\eqref{eq:rhoOfRPhi} and \eqref{eq:thetaOfRPhi} with $r=r_b$ and, in the disk-centered system, corresponds to the polar angle of the boundary point associated with direction $\phi$ in the defect-centered system.

It remains to specify the polar order parameter, which should account for two distinct features. First, the polar order is high in the passive center and close to the disk boundary, but low in an annulus between these two regions, Fig.~\ref{fig:heterogeneousPolarizationAndFlow}(a). Second, polar order decreases when approaching a defect. We use 
\begin{align}
|\mathbf{p}|=p_0(r)(1-\exp\left\{-\rho/\varepsilon\right\}).
\label{eq:ansatzP}
\end{align} 
Here, the second factor approximates the polar order parameter around a defect. The characteristic length $\varepsilon$ is thus determined by the elastic parameters $K$ and $\chi$, Eq.~\eqref{eq:elastic_energy}, $\varepsilon^2 \sim K/\chi$. The factor $p_0$ captures how activity modulates the order parameter.
Explicitly, we write 
\begin{align}
    p_0 &= \bar{p}_0 +\frac{\Delta P_1}{1 + \exp[-(r-(r_i-r_1))/\lambda_1]} \nonumber\\
        &\quad\quad+ \frac{\Delta P_2}{1 + \exp[-(r-(r_b-r_2))/\lambda_2]}.    
\end{align}
For $\Delta P_1 = \Delta P_2 = 0$, the profile is constant, and the parameter $\bar{p}_0$ sets the steady-state value in a passive system. The two additional terms are step functions that enable spatial modulation of the order parameter at the center, near the disk boundary, and in the intermediate region. The first term modifies the polar order parameter outside the passive central region, $r > r_i - r_1$, while the second acts closer to the boundary, $r > r_b - r_2$. These contributions adjust the polar order by $\Delta P_1$ and $\Delta P_2$, respectively. The parameters $r_{1,2}$ control the sizes of the central and boundary regions, and $\lambda_{1,2}$ determine the sharpness of the transitions between them.

Our \textit{ansatz} for the polarization field is qualitatively in good agreement with the numerical solution, Fig.~\ref{fig:heterogeneousPolarizationAndFlow}(c). In particular, it captures the bend and splay distortions in the front and the back of the defect. Other functional forms or a larger number of fit parameters could be used to get a better quantitative agreement between the \textit{ansatz} and the real solution. However, the main goal of the present analysis is to provide qualitative insight into defect motion rather than a quantitative fit to the numerical solution.  

Given the polarization field, which includes in particular the radial defect position, we obtain the flow field from force balance, Eq.~\eqref{eq:forceBalance}. It is equivalent to the Stokes equation, with a force density $\boldsymbol{\nabla}\cdot(\sigma-\sigma^v)$ and can be solved using the corresponding Green's function, App.~\ref{app:StokesFlow}. It turns out that the force density is dominated by the reactive and active components, such that we only use $\boldsymbol{\nabla}\cdot\left(\sigma^a+\sigma^r\right)$ in the following. Whereas the contribution of $\sigma^a$ can be readily obtained analytically, the computation of the contribution of $\sigma^r$ is more involved. We therefore compute both numerically and note merely that the analytics shows that the active force $\boldsymbol{\nabla}\cdot\sigma^a$ is directed \birte{parallel or antiparallel to} $\left(\cos(2\psi_b), \sin(2\psi_b)\right)$. Numerically, we find the same to be true for the contribution from the reactive stress.

Visual inspection indicates that the \textit{ansatz} for the polarization field, combined with the approximation that retains only the reactive and active stress contributions, provides a reasonable estimate of the flow field, Fig.~\ref{fig:heterogeneousPolarizationAndFlow}(d). The main difference compared to the full solution for the velocity field is that the minimum of the approximate velocity occurs farther from the disk center.

\subsection{Azimuthal motion of the defect}

For steady defect motion around the disk, we observe a rigid body like rotation of the fields, see Fig.~\ref{fig:heterogeneousPolarizationAndFlow}. This implies that the system has a constant, non-zero angular momentum given by $L = \int \mathbf{r}\times \mathbf{v}dA$. Note, that the total angular momentum, which includes components resulting from the polarization field, vanishes. The non-zero angular momentum $L$ must arise from a net torque on the fluid, which can be written as $T = \int \mathbf{r}\times\mathbf{f}dA$, where $\mathbf{f}$ are the forces driving the fluid motion. As outlined above, the force density is dominated by the reactive and active components, hence we can estimate the total torque on the system from $T = \int \mathbf{r}\times\boldsymbol{\nabla}\cdot\left(\sigma^a+\sigma^r\right)dA$. Using our ansatz, we numerically estimate the total torque on the system and get an estimate for $\omega\propto T$, 
Fig.~\ref{fig:controlAngularSpeed}(a). This shows a $\sin(2\psi_b)$ relationship that agrees well with the numerical results, Fig.~\ref{fig:controlAngularSpeed}(a).
On the contrary, the numerical results show that varying the boundary angle has a much smaller effect on the radial position of the defect, Fig.~\ref{fig:controlAngularSpeed}(b). 

We can leverage the dependence of the motion of the defect on the boundary angle $\psi_b$ to dynamically control the defect's position in the disk. By alternating $\psi_b$ between $\pm\pi/4$, we cause the defect to propagate back and forth azimuthally at maximal speed and constant radius, resulting in an arc, Fig.~\ref{fig:controlAngularSpeed}(c). The length of the individual arcs can be controlled by adjusting the timing over which $\psi_b$ is varied; the detailed time course of $\psi_b$ for Fig.~\ref{fig:controlAngularSpeed}(e) is given in Tab.~\ref{tab:param_reversing}.

We can also utilize the dependence of the radial position of the defect on $\psi_b$ to generate a 2d path. For example, if we periodically switch $\psi_b$ between $(0.45\pm0.25)\pi$, Tab.~\ref{tab:param_fig8}, we introduce a radial component of $\mathbf{f}$. This leads to the defect tracing a figure eight pattern and shows the potential of this technique, Fig.~\ref{fig:controlAngularSpeed}(d).

\subsection{Controlling the radial defect position}

\begin{figure}[t]
    \centering
    \includegraphics[width=\columnwidth]{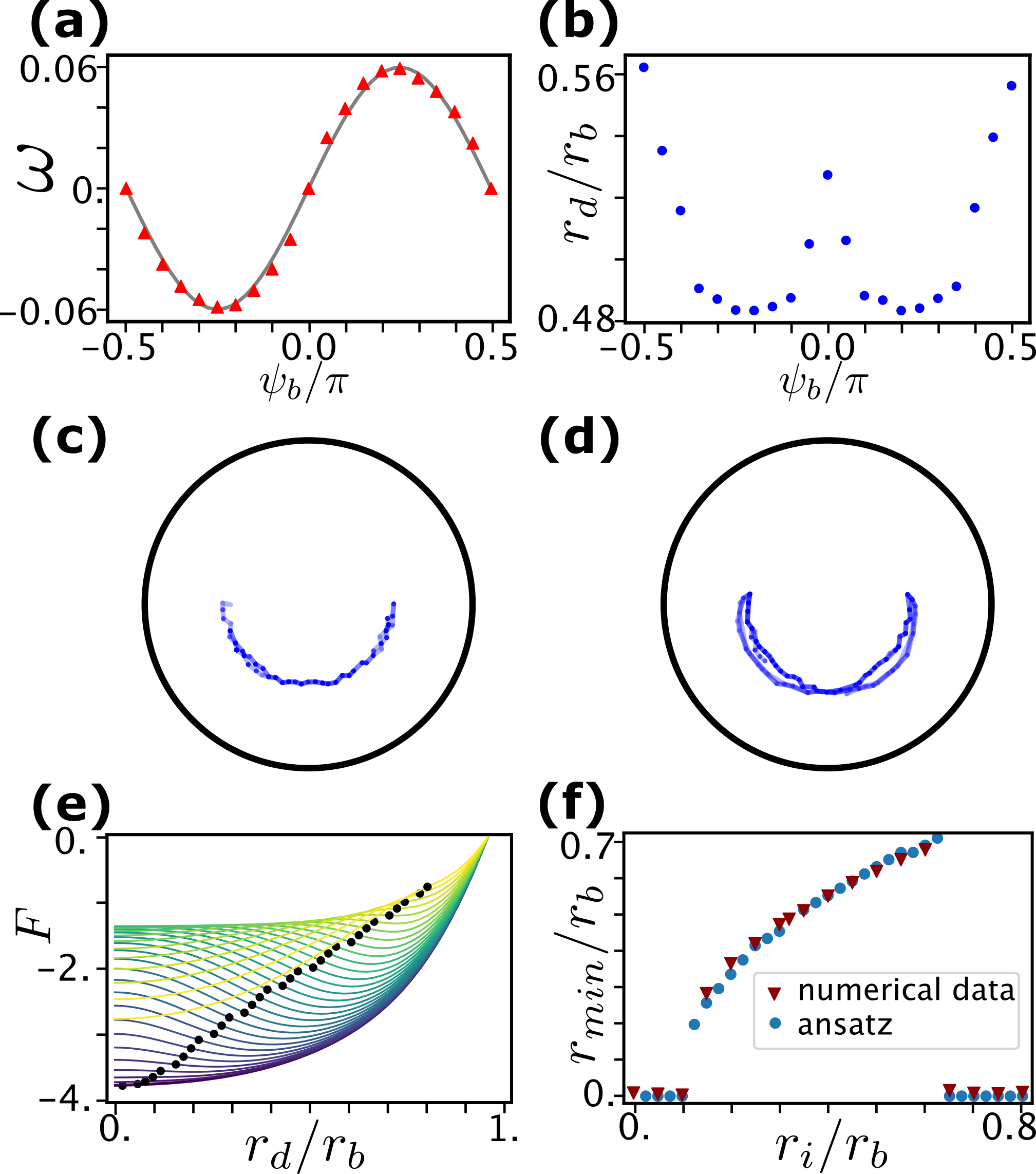}
    \caption{Controlling defect motion through the boundary angle. 
    \textbf{(a)} Angular defect speed $\omega$ as a function of the boundary angle $\psi_b$. Red triangles: numerics, gray line: from \textit{ansatz}. \textbf{(b)} Radial defect position $r_d$ as a function of $\psi_b$. \textbf{(c)} Defect trajectory for $\psi_b$ changing according to Tab.~\ref{tab:param_reversing}. \textbf{(d)} Defect trajectory for $\psi_b$ changing according to Tab.~\ref{tab:param_fig8}.
    \textbf{(e)} Free energy $F$ as a function of the radial defect position $r_d$ for different radii of the passive region $r_i/r_b\in \{0., 0.8\}$ indicated by black dots using the \textit{ansatz} for the polarization field. 
       \textbf{(f)} Position of the elastic potential's minimum as a function of $r_i$. Red triangles: numerics, blue dots: \textit{ansatz}. Parameters as in Tab.~\ref{tab:parameters_constitutiveequ} and $\bar{\lambda}=-0.857$
     }
    \label{fig:controlAngularSpeed}
\end{figure}

Using only the boundary angle $\psi_b$ to guide the defect has two key weaknesses. First, the azimuthal motion is coupled to the radial position, thus one cannot control them independently. Second, we can only access a narrow range of radii, Fig.~\ref{fig:controlAngularSpeed}(b). 

We now show that the defect's radial position is determined predominantly through elastic interactions which are independent of $\psi_b$. To this end, we evaluate the free energy, Eq.~\eqref{eq:elastic_energy} for the \textit{ansatz} polarization field \eqref{eq:ansatzP}. Note that through the dependence on the radius $r_i$ of the passive central disk as well as on $\Delta P_{1,2}$, the results based on the \textit{ansatz} also depend on activity. 

In Figure~\ref{fig:controlAngularSpeed}(e), we show the free energy as a function of the defect position $r_d$ for different values of the radius $r_i$. For sufficiently small values of $r_i$, the energy exhibits a single minimum at $r_{\mathrm{min}} = r_d$. When $r_i$ falls below a critical value, this minimum is localized at the disk center, $r_{\mathrm{min}} = 0$. Above the critical value, $r_{\mathrm{min}}$ increases monotonically with $r_i$, Fig.~\ref{fig:controlAngularSpeed}(f). Within our numerical accuracy, the transition appears to be continuous. 

The radial defect positions obtained from our \textit{ansatz} agree very well with the numerical solutions, Fig.~\ref{fig:controlAngularSpeed}(f). This leads to an intuitive picture of the underlying physical mechanism. Within the \textit{ansatz}, the value of $r_\textrm{min}$ is set by the competition between two effects. First, since the polar order is small at the defect but fixed to unity at the boundary, gradient contributions to the free energy increases as the defect is positioned closer to the boundary. This effectively introduces a repulsive force between the boundary and the defect. defect but fixed to unity at the boundary. As a consequence, the defect is pushed towards the disk center. 
Second, the term proportional to $\chi$ in the free energy becomes important, when the defect is located in the passive region.  
As soon as the passive region is large enough, the gain in energy by moving the defect further to the center and thus reducing the gradients in $\mathbf{p}$ is outweighed by the increase in energy due to the deviation of the order parameter from its preferred value $P_0$. Balancing both contributions leads to $r_d>0$ for sufficiently large $r_i$. 

As the size of the passive disk reaches a second critical value, the energy is essentially the same as for a homogeneous passive system as long as $r_d$ is sufficiently small. As a consequence, the energy exhibits a second minimum at $r_d = 0$ in this case. While this suggests coexistence of two steady states, we only observe the global minimum numerically.

\subsection{Independent control of angular speed and radius}

As we have seen, the boundary angle $\psi_b$ can be used to set the angular defect velocity and the radial defect position. To reach independent control of both quantities, we utilize the radius $r_i$, which only affects the radial defect position. Specifically, we 
chose to implement a feedback controller that modulates $r_i$ over time to target a specific radial position of the defect. 

\begin{figure}[t]
    \centering
    \includegraphics[width=0.9\columnwidth]{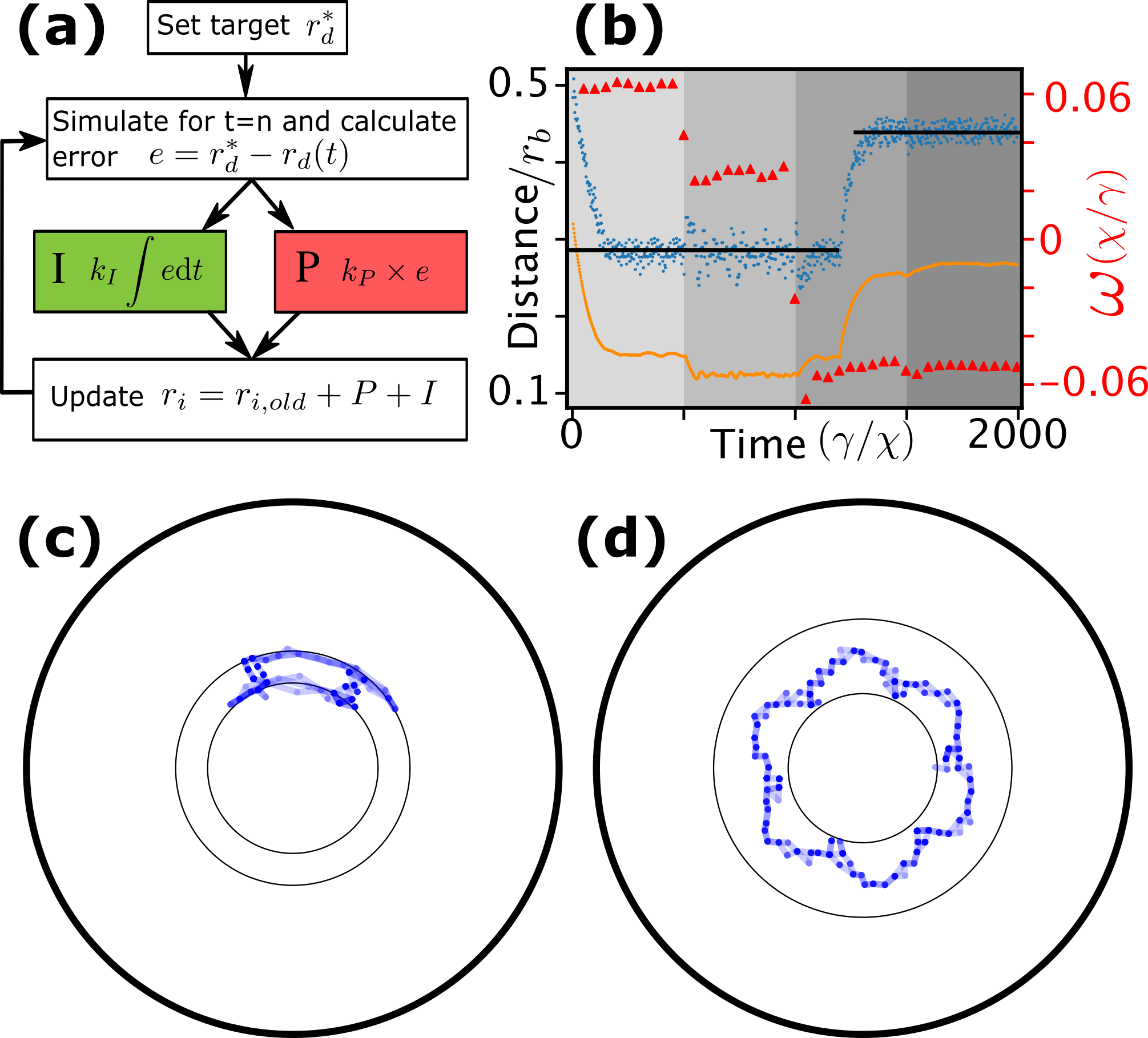}
    \caption{Dynamic feedback control of the defect location.
    \textbf{(a)} Schematic of the PI-Controller.
    The controller maintains the radial distance of the defect at a desired setpoint $r_d^*$ by dynamically adjusting the radius of the passive core, $r_i$. For each $r_i$, the system is simulated for $\text{n}=10$ time steps to determine the resulting $r_d$, which yields the error that is fed back into the controller. The new $r_i$ is calculated as $r_i=r_{i, old}+P+I$, where the $P$ denotes the proportional part and $I$ the integral part of the controller. 
    \textbf{(b)} Independent control of defect radius $r_d$ and defect angular velocity $\omega$. The radial setpoint changes from $r_d^*/r_b=7/25$ to $r_d^*/r_b=11/25$ at time step $1200$ (black lines). Meanwhile, the boundary angle $\psi_b$ increases each $500\gamma/\chi$ by $\pi/5$ starting from  $\psi_b=\pi/4$. Blue dots: actual radial defect position, red triangles: angular velocity, orange dots: $r_i$     Parameter values as in Tab.~\ref{tab:parameters_constitutiveequ} and $\bar{\lambda}=-0.857$
    \textbf{(c)} Defect trajectory (blue dots) when prescribing a closed loop with right angle corners. Static parameter values as in \birte{Tab.\ref{tab:parameters_constitutiveequ}} and $\bar{\lambda}=-0.857$, dynamic parameters: Tab.~\ref{tab:param:loop}. Control parameters: $k_P=5\cdot10^{-4}$ and $k_I=5\cdot 10^{-3}$.
    \textbf{(d)} Defect trajectory (blue dots) when prescribing a hanagata shape with internal angles of $\pi$. Static parameter values as in \birte{Tab.~\ref{tab:parameters_constitutiveequ}} and $\bar{\lambda}=-0.514$, dynamic parameters: Tab.~\ref{tab:hanagata}. 
    In (c) and (d), gray circles correspond to the extreme values of $r_d^*$. }
    \label{fig:PIcontroller}
\end{figure}

We opt for a proportional--integral controller (PI-Controller)~\cite{Geering:2013}. We define a target radius for the defect, here denoted $r^*_d$. The error in the target observable is calculated as $r_d-r_d^*$, and the control parameter $r_i$ is updated proportional to the instantaneous  and the time-integrated  error, Fig.~\ref{fig:PIcontroller}(a). This approach allows us to avoid asymptotic deviations from the target value at steady state~\cite{Geering:2013}. The parameters of the PI-controller must be carefully tuned to achieve the desired control, App.~\ref{app:pi_control} that is robust to changes in activity, \red{Fig.~\ref{fig:PIControl}}.

In Figure~\ref{fig:PIcontroller}(b), we demonstrate the efficacy of controlling the radial position of the defect with the PI controller. 
Changes in the radial position $r_d$ following changes in $\psi_b$ are rapidly compensated by the PI controller, Fig.~\ref{fig:PIcontroller}(b). When changing the target radius $r_d^*$, the direction of circular defect motion remains unchanged whereas its speed changes. This demonstrates that we can independently control the defect's radial position $r_d$ and its angular speed by using the PI controller and adjusting the boundary angle $\psi_b$. 

The enhanced control of the defect made possible by the PI controller allows us to generate more complex patterns that were previously not possible. We demonstrate this by generating a closed loop with right angle corners, Fig.~\ref{fig:PIcontroller}(c), Tab. A~\ref{tab:param:loop}, and a hanagata shape, Fig.~\ref{fig:PIcontroller}(d), Tab. A~\ref{tab:hanagata}. Even more complex patterns would be possible using appropriate sequences of target radii and boundary conditions. 

\section{Discussion}

In summary, we investigated the behavior of active polar fluids confined to a 2D disk with a single topologically constrained $+1$ defect. For a heterogeneous activity pattern, the defect moves in a circle around the center.  
The circular motion depends on the boundary angle of the polar field and the size of the active annulus.
Based on these dependencies we imposed target trajectories on the defect by means of a PI Controller.

Combining heterogeneous activity patterns with changes of the boundary conditions gives ample possibilities for controlling the motion of individual defects. Heterogeneous activity patterns different from the one studied in this work have been explored theoretically before. On a disk, complex activity patterns in combination with imposing vorticity were shown to set the sense of rotation of a pair of $+1/2$ defects~\cite{Norton.2020}. Elementary activity patterns were identified to create, move, and braid $1/2$ defects~\cite{Shankar.2024}. These included $-1/2$ defects, which do not self-propel. By continuously updating activity patterns $+1$ defects, which do not self-propel either were moved in a polar Toner-Tu system~\cite{Ghosh.2024}.

The effects on the orientation of $+1/2$ defects through spatial jumps in activity similar to the border between the passive inner circle and the active annulus considered in the present work were explored in Ref.~\cite{ronning_defect_2023}. Reorientation of $+1/2$ defects were shown to be feasible by alignment with activity gradients~\cite{Tang.2021, Ruske.2022}.

Experimentally, defect control in active systems has been achieved in reconstituted motor-filament systems~\cite{Zhang.2021,Nishiyama.2025}. In these studies, light was used to increase motor activity in certain regions, which led to an accumulation of +1/2 defects and allowed the authors to bias the motion of single $+1/2$ defects along coarsely predefined paths~\cite{Zhang.2021}. A PI-controller was implemented to control the coarse-grained flow speed of a microtubule-based nematic~\cite{Nishiyama.2025}. Other control mechanisms could be envisioned, for example, based on reinforcement learning as was used in a theoretical study to control $\pm1/2$ defects~\cite{floydTailoringInteractionsActive2025}.

The potential advantage of the system introduced in the present work is a higher precision of the control of defect trajectories, which relies on the ability to set the boundary angle $\psi_b$. In reconstituted motor-filament systems, this might be challenging, although the capacity of magnetic fields to align microtubules when mixed with an appropriate synthetic nematic liquid crystal has been demonstrated experimentally~\cite{Guillamat.2016}. For controlling cell monolayers, patterned surfaces provide promising alternatives~\cite{Guillamat.2022, zhaoIntegerTopologicalDefects2025,coyleCellAlignmentModulated2022, endresenTopologicalDefectsInteger2021}. While currently, most patterns are static, new developments aim also at dynamic surface patterning~\cite{Isomaeki.2022}. Given the importance of topological defects in tissue morphogenesis~\cite{Maroudas-Sacks.2021, Guillamat.2022, Ravichandran.2025}, we consider this a particularly interesting research direction to follow.

\begin{acknowledgments}
BCG thanks FOSTER (STUDENTRESEARCH@TUD) for funding a student research stay in Geneva in 2023.
DJGP gratefully acknowledges funding from the Swiss National Science Foundation (SNSF Starting Grant TMSGI2\_211367)
\end{acknowledgments}

\appendix
\section{Numerics}
\label{app:numerics}

\subsection{Solution algorithm}

Our custom simulation code is written in C++ using the OpenFPM library for scalable scientific computing~\cite{Incardona.2018}. For the discretization of the differential operators, Discretization-Corrected Particle Strength Exchange (DC-PSE)~\cite{Schrader.2010}\cite{singhMeshfreeCollocationScheme2023a} is used, which is implemented in OpenFPM via a template-expression system for partial differential equations~\cite{Singh.2021}.

DC-PSE is a meshfree collocation method, and discretization points do not need to lie on a grid.
A (boxsize $\times$ boxsize) Cartesian grid is created as a template for the particles of DC-PSE, where $\text{boxsize}=2 r_b$, but can be changed independently if higher resolution is wanted. 
A circular grid is added, whose center is defined at (boxsize/2, boxsize/2).
The bulk particles are initially placed on a Cartesian grid and retained only if their positions lie within a radius, $r_b=25-7\text{spacing}/6$ from the domain center. To avoid directional bias introduced by the underlying Cartesian grid, a uniformly distributed random perturbation in the interval $(-0.2/r_b, 0.2/r_b)$ is added independently to each coordinate. Without this perturbation, the discretization of the differential operators in the framework tends to favor gradients aligned with the Cartesian axes, producing nonphysical quadrangular flow patterns.

Additionally, particles near the domain boundary are classified as boundary particles at radius $r_b- 3\text{spacing}/4$. A corresponding ghost-particle layer is generated on the circular boundary at radius $r_b$ 
by radially projecting nearby particles onto the outer boundary. These layers are used to impose Dirichlet anchoring conditions on the polarization field.
At the boundary and ghost particles, the polarization is fixed to unit magnitude $|\mathbf{p}|^2=1$ and to a constant angle $\psi_b$, where $\psi$ is defined with respect to the radial direction in the disk-centered coordinate system. 
No-slip boundary conditions are imposed independently on the velocity field through the Stokes solver.
The characteristic length scale for the polarity magnitude to go from 1 to 0 can be determined as $\varepsilon=\sqrt{K/\chi}=\sqrt{10}$, see Tab.~\ref{tab:parameters_constitutiveequ} for the physical parameters. To avoid finite-size effects, the radius of the disk $r_b$ has to be chosen such that $r_b>>\varepsilon$. Furthermore, the interaction cutoff radius $\text{rCut}$, which determines the neighborhood size used in the differential operator stencil, must resolve the polarization healing region. Therefore,
$\text{rCut}\geq \varepsilon \times\text{spacing}$, corresponding to approximately $\sqrt{10} \times\text{spacing}$. 

The polarity field is first relaxed to a stationary bulk configuration under the imposed boundary condition assuming vanishing flow. The resulting configuration is used as an initial condition for further time integration, which is realized by using the adaptive Adams--Bashforth--Moulton method. This is a multi-step method combining an explicit predictor (Adams-Bashforth) with an implicit corrector (Adams-Moulton) step \cite{singh_integrating_2026}, from the {\tt boost::numerics::odeint} library using their {\tt make::controlled option}. The time tolerance and the relative time tolerance are both set to $7e-3$.  This is combined with the OpenFPM methods {\tt DCPSE\_scheme} for discretizing spatial derivatives and {\tt petsc\_solver} for solving the Stokes problem.
The incompressibility constraint $\partial_{\gamma} v_{\gamma}=0$ is imposed by using the pressure-correction algorithm as previously described \cite{Singh:2023c} at each simulation step. The velocity field itself is generated self-consistently through the active stresses arising from the polarization field and the no-slip hydrodynamic boundary conditions.

Defect tracking, output writing, and the PI-Controller, as well as changes in the activity are implemented as {\tt Observer} methods to the Adams--Bashforth--Moulton method. The defect is tracked as the local minimum in polarization magnitude, which is broadcast and compared across all compute processes to find the defect position for controller evaluation.

\subsection{System parameters}
As introduced before in Sec.\ref{subsec:hydrodynamics_of_an_active_polar_fluid}, we use the energy scale $K_S$, the length scale $\sqrt{K_S/\chi}$ and the time scale $\gamma/\chi$. We get the two activities $\bar{\lambda}=\frac{\gamma\lambda\Delta\mu}{\chi}$ and $\bar{\zeta}=\frac{\zeta\Delta\mu}{\chi}$ as well as the dimensionless viscosity $\bar{\eta}=\eta/\gamma$.
We introduce the dimensionless Frank constant $\overline{dK} =\frac{K_B-K_S} {K_S}$, which captures whether the system favors bend or splay deformations in $\mathbf{p}$.
Table \ref{tab:parameters_constitutiveequ} shows the values for the system parameters used in numerical simulations if not stated otherwise.

\begin{table}[htb]
    \centering
    \begin{tabular}{ |c | c |}
        \hline
        $\bar{\eta}$& $2$ \\
        \hline
        $\gamma/\chi$ & 1 \\
        \hline
        $\sqrt{K_S/\chi}$ & $\sqrt{10}$ \\
        \hline
        $\bar{\lambda}$ & $[-0.9, 0]$ \\
        \hline
        $\bar{\zeta}$ & $-\bar{\lambda}/7.5$ \\
        \hline
        $\overline{K}_B$ & 1 \\
        \hline
        $r_b$ & 25 \\
        \hline
        $r_i$ & $[0, 0.8]r_b$ \\
        \hline
        $\nu$  & $2$ \\
        \hline
    \end{tabular}
    \vspace{3mm}
    \caption{Values of the non-dimensional parameters used in the main text, unless stated otherwise.}
    \label{tab:parameters_constitutiveequ}
\end{table}

\subsection{PI-Controller configuration}\label{app:pi_control}
To achieve stable control, the amplifications $k_P$ and $k_I$ must be chosen such that overshooting, oscillations, and transient errors are balanced. As expected, for $k_P\neq0$ and $k_I=0$ we observe an asymptotic steady-state error known as {\it droop}~\cite{Nishiyama.2025}. Introducing an integral part can reduce droop, but if $k_I$ is too large the system is prone to overshooting. If $k_P$ is too large, the system becomes oscillatory and eventually unstable. Values too small do not achieve the desired control.
Here, we first set $k_I$ to be small and decrease $k_P$  from a large value until oscillations vanish. Then $k_I$ was subsequently increased, until droop was minimized. 
Using these $k_P=5\cdot 10^{-4}$ and $k_I=5\cdot 10^{-3}$, we achieve stable control of $r_d$. 
Additionally, an anti-windup mechanism is introduced with limits $\pm 2\text{spacing}$.

\subsubsection{Independence of radial control and activity.}

\begin{figure}[h]
    \centering
            \includegraphics[width=0.8\columnwidth]{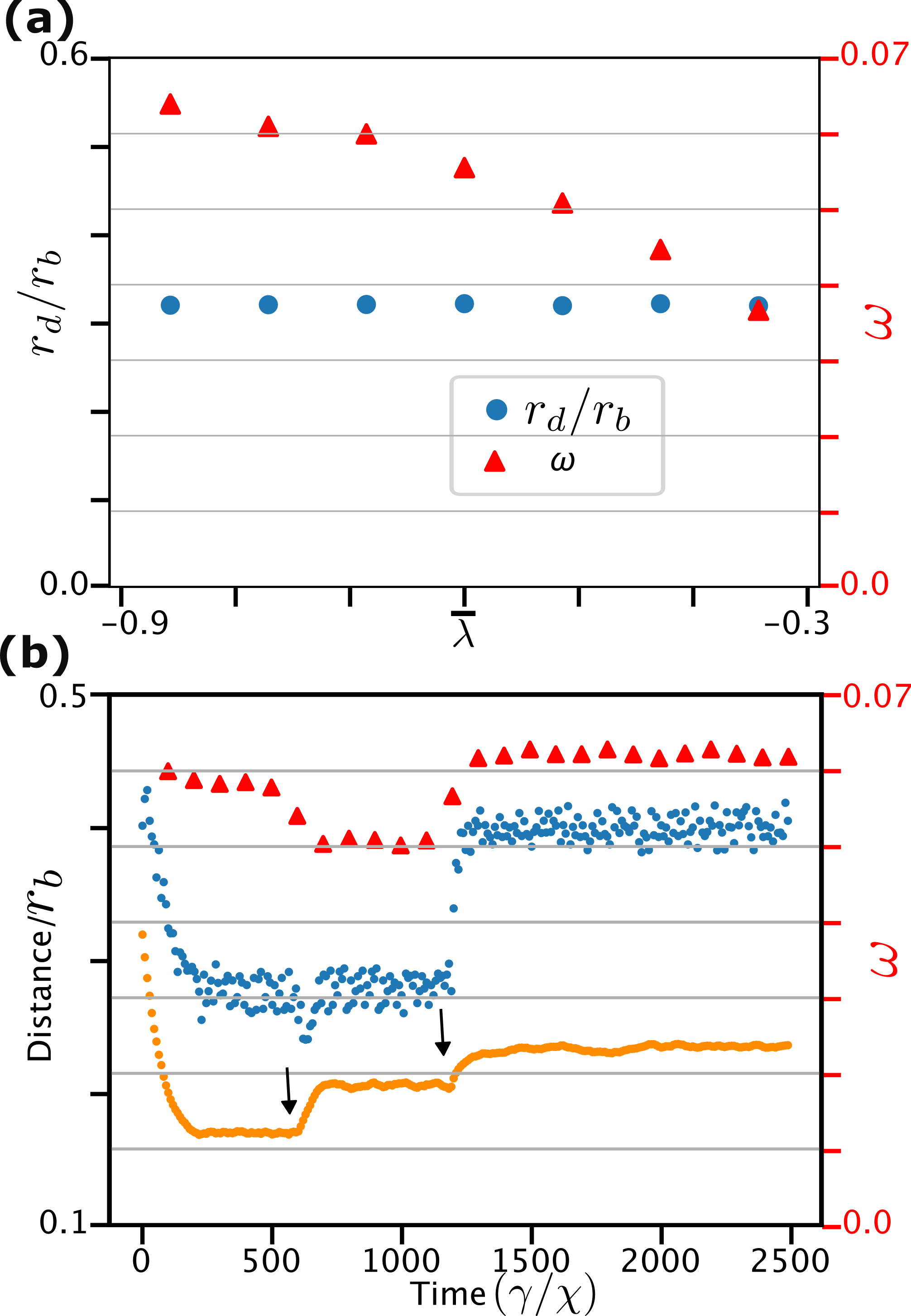}
            \label{fig:w_over_act_const_rd}
    \caption{\textbf{(a)} Angular velocity of the defect for varying activity while holding the radial position of the defect constant with the PI controller setpoint $r_d^*/r_b=8/25$.
     \textbf{(b)} Controlled dynamics under abrupt change of $r_d^*/r_b$ from $7/25$ to $10/25$ at time $1200\gamma/\chi$.
    At times $600\gamma/\chi$ and $1200\gamma/\chi$, 
    the activity $\bar{\lambda}$ is changed, first from $-0.686$ to $-0.514$, then to $-0.857$\birte{, indicated by black arrows}. This changes the angular defect velocity (red triangles, right scale). Throughout, the controller maintains the defect position near the setpoint.}
    \label{fig:PIControl}
\end{figure}

Fig.~\ref{fig:PIControl}a shows \birte{that for a range of activities, the defect position can be held at a constant $r_d/r_b=8/25$ by the PI-Controller. Further, we can observe} that the angular velocity of the defect reduces as activity is reduced for a constant $r_d$. 

Fig.~\ref{fig:PIControl}b shows that the PI controller is able to maintain the defect's radial position during changes in the activity. The activity is reduced at time $600 \gamma/\chi$, which prompts the PI-controller to increase $r_i$ to maintain $r_d$. When activity is reduced, the difference in the stable magnitude of the polar field between the active and passive areas is reduced, which in turn reduces the gradient of the elastic potential of the defect between these two regions, thus $r_i$ must increase to compensate. At time $1200\gamma/\chi$, the setpoint $r_d^*$ as well as the activity are changed to $10/25$ and to $\bar{\zeta}=0.114$ respectively. The PI-Controller successfully regulates $r_i$ to reach and stabilize the desired setpoint.

\section{Steady state for homogeneous activity}
\label{app:homogenousActvity}

\begin{figure}[tb]
    \centering
    \includegraphics[width=0.9\linewidth]{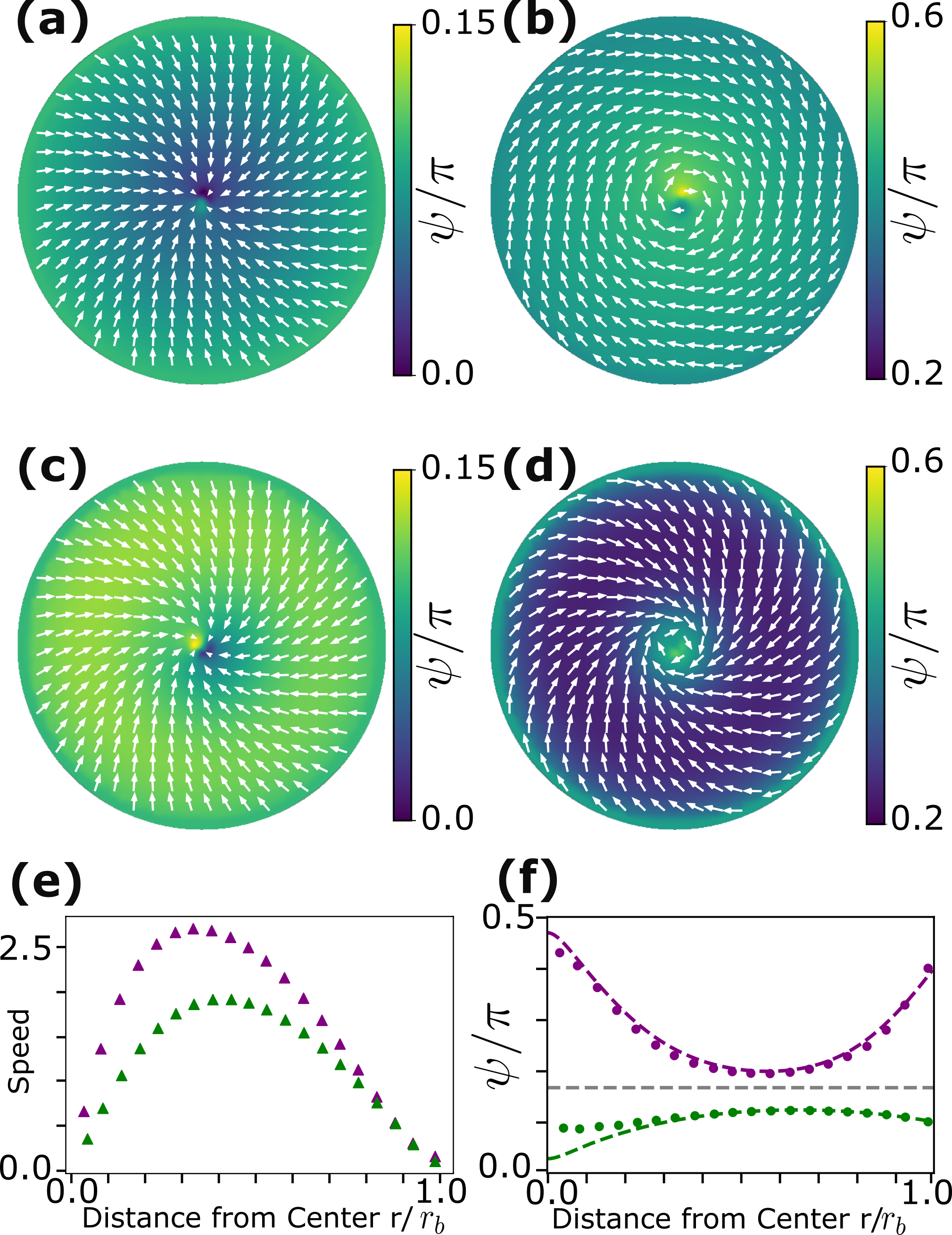}
    \caption{Steady states for $\bar{\lambda}=0$, $\Delta\mu=0.8$. (a-d) Polarization fields for $\overline{dK}=1$, $\bar{\zeta}=0$, and $\psi_b=\pi/10$ (a), $\overline{dK}=-1$, $\bar{\zeta}=0$, and $\psi_b=2\pi/5$ (b), $\overline{dK}=1$, $\bar{\zeta}=-1.6$, and $\psi_b=\pi/10$  (c), and $\overline{dK}=-1$, $\bar{\zeta}=-1.6$, $\psi_b=2\pi/5$ (d). White arrows: $\mathbf{p}$, colors: polarization angle $\psi/\pi$.
    (e) Azimuthal average of the angular speed $\omega$ for (c, green) and (d, purple).
    (f) Azimuthal average of the angle $\psi$ for (c, green) and (d, purple). Green and purple dashed lines: solutions to Eq.~(\ref{eq:nonconst_angle}), gray dashed line: Leslie angle $\psi_L$, Eq.~(\ref{eq:Leslie_angle}).}
    \label{fig:steadyStateHomogenousActivity}
\end{figure}

In this section, we study the base case of homogeneous activity for different values of the elastic anisotropy $d\overline{K}=\overline{K}_B-1$, $\bar{\lambda}=0$, and a constant value of the angle $\psi_b$ along the boundary. The latter condition implies the existence of topological defects with a total charge $+1$, Fig.~\ref{fig:steadyStateHomogenousActivity}.

In the absence of active stress, $\bar{\zeta} = 0$, the system equilibrates to $\mathbf{v}=0$ and a polarization field $\mathbf{p}$ with $|\mathbf{p}|^2=1$. There is one defect of $\mathbf{p}$ situated in the disk center. The orientation of the equilibrium polarization field is governed by $\overline{dK}$. When bend deformations are energetically more costly than splay, $\overline{dK} > 0$, the angle $\psi$ decreases monotonically toward zero at the disk center, corresponding to an aster defect, Fig.~\ref{fig:steadyStateHomogenousActivity}(a). Conversely, when bend is energetically favored, $\overline{dK} < 0$, $\psi$ approaches $\pm\pi/2$ at the center, yielding a vortex defect, Fig.~\ref{fig:steadyStateHomogenousActivity}(b). Since the imposed boundary angle differs from these values, the steady state polarization field smoothly transitions from an aster or vortex defect in the center to a spiral at the boundary. 

For a contractile activity $\bar{\zeta}=-1.6$, the steady state retains its rotational symmetry, Fig.~\ref{fig:steadyStateHomogenousActivity}(c,d). In contrast to the equilibrium case, however, it exhibits an azimuthal flow, Fig.~\ref{fig:steadyStateHomogenousActivity}(e). The flow is driven by gradients in the active stress resulting from the radial gradient in $\psi$. The flow alignment term in Eq.~\ref{eq:time_evolution_p} leads to a feedback between the flow and the angle $\psi$ that eventually balances and leads to a steady profile in which $\psi$ varies non-monotonically, Fig.~\ref{fig:steadyStateHomogenousActivity}(f).

We compare our results obtained by numerically solving the dynamic equations \eqref{eq:time_evolution_p} and \eqref{eq:forceBalance} to solutions of the steady state equation for the angular component of the polarization, when assuming rotational invariance and $|\mathbf{p}|=1$. Under these assumptions, it reads~\cite{Kruse:2004il}
\begin{align}
\label{eq:nonconst_angle}
		h_{\perp}[4\bar{\eta} + (\nu^2 + 1 - 2 \nu \cos{2\psi})] = &\nonumber\\
        &\hspace{-1cm} \sin{2\psi}(\nu \cos{2 \psi} -1) (\bar{\zeta}+\nu\bar{\lambda}),
\end{align}
where
\begin{align}
        h_{\perp}&=(1+\overline{dK} \cos^2{\psi})\left[\psi'' + \frac{\psi'}{r}\right] - \frac{\overline{dK}}{2}\sin{2\psi}\left[\frac{1}{r^2} + \psi'^2 \right].
\end{align}
Here, $\psi=\psi(r, t)$ and $\psi'=\partial_r\psi$.
In the case $\overline{dK}=0$ and assuming free anchoring at the boundary, such that $\psi_b$ adapts to the bulk, $h_\perp=0$ and the polarization assumes the Leslie angle $\psi_L$ with 
\begin{align} 
\label{eq:Leslie_angle}
	\cos (2 \psi_L) = \frac{1}{\nu},
\end{align}
if $|\nu|\ge1$. In the case of strong anchoring, that is, when imposing $\psi_b$ and $\partial_r \psi(0, t)=0$, the numerical solution of Eq.~\eqref{eq:nonconst_angle} agrees well with the steady state solution obtained by solving the full dynamic equations, Fig.~\ref{fig:steadyStateHomogenousActivity}(f). Far from the boundaries, the polarization angle $\psi$ approaches the Leslie angle $\psi_L$.

\section{Stokes flow for an arbitrary force field}
\label{app:StokesFlow}

To compute the flow for a given polar texture, we follow a methodology similar to Ref.~\cite{giomi2014defect}. At low Reynolds number, the flow field $\mathbf{v}$ satisfies the Stokes equation 
\begin{align}
\eta\Delta \mathbf{v} &= \mathbf{f},
\end{align}
where $\mathbf{f}$ is in our case given by $\boldsymbol{\nabla}\cdot\left(\sigma^a+\sigma^r\right)$. 

We compute the force density $\mathbf{f}$ numerically on a 256$\times$256 grid. The flow field is then given by the convolution of the two-dimensional Oseen tensor with the force density
\begin{equation}
    v_\alpha(\mathbf{r}) = \int\int G_{\alpha\beta}(\mathbf{r} - \mathbf{r}')f_\beta(\mathbf{r}')\, dA',
\end{equation}
where the two-dimensional Oseen tensor is
\begin{equation}
    G_{\alpha\beta}(\underline{r}) = \frac{1}{4\pi\eta}\left[ [\log(\mathcal{L}/r) - 1]\delta_{\alpha\beta} + \frac{r_\alpha r_\beta}{r^2}\right]\, .
\end{equation}
Here, $\mathcal{L}$ is a length scale that can be adjusted to give the correct boundary behavior. This is further augmented by a homogeneous solution to the Stokes equation to satisfy the no-slip boundary condition~\cite{di2008hydrodynamic}.

\FloatBarrier
\section{Instruction sets for complex defect paths}\label{appsubsec:complex_control}

Below we include the precise changes in boundary angle (and target radius in the case where a PI controller was used) to generate the complex defect paths shown in the main text.

\begin{table}
    \begin{tabular}{c|c}
        time (s) & $\psi_b$ \\
        \hline
        0-50 & $0.25\pi$ \\
        50-100 & $0\pi$ \\
        100-150 & $-0.25\pi$ \\
        150-200 & $0\pi$ \\
        200-250 & $0.25\pi$
    \end{tabular}
    \caption{Timing and values of changes in boundary angle to generate Fig.~\ref{fig:controlAngularSpeed}c, $\bar{\zeta} = 0.114$.}
    \label{tab:param_reversing}
\end{table}

\begin{table}
    \begin{tabular}{c|c}
        time (s) & $\psi_b$ \\
        \hline
        0-20 & $0.2\pi$ \\
        20-60 & $0.7\pi$ \\
        60-100 & $0.2\pi$ \\
        100-140 & $0.7\pi$ \\
        140-180 & $0.2\pi$ \\
        180-220 & $0.7\pi$ \\
        220-160 & $0.2\pi$ \\
    \end{tabular}
    \caption{Timing and values of changes in boundary angle to generate  Fig.~\ref{fig:controlAngularSpeed}d, $\bar{\zeta} = 0.114$.}
    \label{tab:param_fig8}
\end{table}

\begin{table}
    \begin{tabular}{c|c|c}
        time (s) & $\psi_b$ &$r_d^*/r_b$\\
        \hline
        0-20 & $0.0\pi$ & 0.44\\%change psi
        20-40 & $0.13\pi$ & 0.44\\%change psi
        40-80 & $0\pi$ & 0.44\\%change r_d*
        80-420 & $0\pi$ & 0.32\\%change psi
        420-440 & $-0.13\pi$ & 0.32\\%change psi
        440-480 & $0.\pi$ & 0.32\\%change r_d*
        480-620 & $0.\pi$ & 0.44 \\%change psi
        620-640 & $0.13\pi$ & 0.44\\%change psi
        640-685 & $0.\pi$ & 0.44\\%change r_d*
        685-820 & $0.\pi$ & 0.32\\%change psi
        820-840 & $-0.13\pi$ & 0.32\\%change psi
        840-885 & $0.\pi$ & 0.32\\%change r_d*
        885-1020 & $0.\pi$ & 0.44\\%change psi
        1020-1040 & $0.13\pi$ & 0.44\\%change psi
        1040-1090 & $0.\pi$ & 0.44\\%change r_d*
        1090-1220 & $0.\pi$ & 0.32\\%change psi
        1220-1240 & $-0.13\pi$ & 0.32\\%change psi
        1240-1400 & $0.\pi$ & 0.32\\
    \end{tabular}
    \caption{Timing and values of changes in boundary angle and target radius to generate Fig.~\ref{fig:PIcontroller}c, $\bar{\zeta} = 0.114$.}
    \label{tab:param:loop}
\end{table}

\begin{table}[]
    \centering
    \begin{tabular}{c|c|c||c|c|c}
        time (s) & $\psi_b$ & $r_d^*$ & time (s) & $\psi_b$ & $r_d^*/r_b$\\
        \hline
        0--60     & $0.016\pi$     & 0.56   & 1080--1140 & $0.016\pi$     & 0.56 \\
        60--120   & $0.016\pi$     & 0.28   & 1140--1200 & $0.016\pi$     & 0.28 \\
        120--180  & $0$            & 0.28   & 1200--1260 & $0$            & 0.28 \\
        180--240  & $0.016\pi$     & 0.56   & 1260--1320 & $0.016\pi$     & 0.56 \\
        240--300  & $0.016\pi$     & 0.28   & 1320--1380 & $0.016\pi$     & 0.28 \\
        300--360  & $0$            & 0.28   & 1380--1440 & $0$            & 0.28 \\
        360--420  & $0.016\pi$     & 0.56   & 1440--1500 & $0.016\pi$     & 0.56 \\
        420--480  & $0.016\pi$     & 0.28   & 1500--1560 & $0.016\pi$     & 0.28 \\
        480--540  & $0$            & 0.28   & 1560--1620 & $0$            & 0.28 \\
        540--600  & $0.016\pi$     & 0.56   & 1620--1680 & $0.016\pi$     & 0.56 \\
        600--660  & $0.016\pi$     & 0.28   & 1680--1740 & $0.016\pi$     & 0.28 \\
        660--720  & $0$            & 0.28   & 1740--1800 & $0$            & 0.28 \\
        720--780  & $0.016\pi$     & 0.56   & 1800--1860 & $0.016\pi$     & 0.56 \\
        780--840  & $0.016\pi$     & 0.28   & 1860--1920 & $0.016\pi$     & 0.28 \\
        840--900  & $0$            & 0.28   & 1920--1980 & $0$            & 0.28 \\
        900--960  & $0.016\pi$     & 0.56   & 1980--2040 & $0.016\pi$     & 0.56 \\
        960--1020 & $0.016\pi$     & 0.28   & 2040--2100 & $0.016\pi$     & 0.28 \\
        1020--1080& $0$            & 0.28   & 2100--2160 & $0$            & 0.28 \\
    \end{tabular}
    \caption{Timing and values of changes in boundary angle and target radius to generate Fig.~\ref{fig:PIcontroller}d. $\bar{\zeta} = 0.069$.}
    \label{tab:hanagata}
\end{table}
\FloatBarrier
\bibliography{references}

\end{document}